\documentclass[useAMS,usenatbib]{mn2e}
\usepackage{graphicx} \usepackage{rotating}
\usepackage{txfonts}
\usepackage{url} 
\usepackage{fmtcount}
\usepackage{upgreek}

\def\nt/f{Nuclear Technology/Fusion}

\title[First Detection of $^3$He$^+$ in the PN IC\,418]
{\begin{center}
First Detection of $^3$He$^+$ in the Planetary Nebula IC\,418
\end{center}}
\author[L. Guzman-Ramirez et al.]{L. Guzman-Ramirez$^{1,2}$\thanks{E-mail: guzmanl@eso.org}, J.~R. Rizzo$^3$, A.~A. Zijlstra$^{4}$, C. Garc\'{\i}a-Mir\'o$^{5}$, \newauthor C. Morisset$^{6}$ \& M.~D. Gray$^{4}$ \\
$^{1}$European Southern Observatory, Alonso de C\'ordova 3107, Casilla 19001, Santiago, Chile\\
$^{2}$Leiden Observatory, Leiden University, Niels Bohrweg 2, 2333 CA Leiden, The Netherlands\\
$^{3}$Centro de Astrobiolog\'ia (INTA-CSIC), Ctra. M-108, km. 4, 28850, Torrej\'on de Ardoz, Spain\\
$^{4}$Jodrell Bank Centre for Astrophysics, School of Physics and Astronomy, University of Manchester, Manchester, M13 9PL, UK\\
$^{5}$Madrid Deep Space Communications Complex, Ctra.~M-531, km.~7, E-28294 Robledo de Chavela, Madrid, Spain \\
$^{6}$Instituto de Astronom\'ia, Universidad Nacional Aut\'onoma de M\'exico, 04510 M\'exico D.F., M\'exico \\
}

\date{Released 2015 Xxxxx XX}

\pagerange{\pageref{firstpage}--\pageref{lastpage}} \pubyear{2015}

\def\LaTeX{L\kern-.36em\raise.3ex\hbox{a}\kern-.15em
    T\kern-.1667em\lower.7ex\hbox{E}\kern-.125emX}

\begin{document}

\date{Accepted 2016 xxx. Received 2015 xxx; in original form 2015 xxx}

\pagerange{\pageref{firstpage}--\pageref{lastpage}} \pubyear{2015}

\maketitle

\label{firstpage}

\begin{abstract}
The $^3$He isotope is important to many fields of astrophysics,
including stellar evolution, chemical evolution, and cosmology.  The
isotope is produced in low-mass stars which evolve through the
planetary nebula (PN) phase. $^3$He abundances in PNe can
help test models of the chemical evolution of the Galaxy.  We present the detection of the $^3$He$^+$ emission line
using the single dish Deep Space Station 63, towards the PN IC\,418. We
derived a $^3$He/H abundance in the range
1.74$\pm$0.8$\times$10$^{-3}$ to 5.8$\pm$1.7$\times$10$^{-3}$,
depending on whether part of the line arises in an outer ionized halo.
The lower value for $^3$He/H ratio approaches values predicted by
stellar models which include thermohaline mixing, but requires that
large amounts of $^3$He are produced inside low-mass stars which
enrich the interstellar medium (ISM). However, this over-predicts the
$^3$He abundance in H\,{\sc ii} regions, the ISM, and proto-solar
grains, which is known to be of the order of 10$^{-5}$. This discrepancy
questions our understanding of the evolution of the $^3$He, from
circumstellar environments to the ISM.
\end{abstract}

\begin{keywords}
circumstellar matter -- radio: abundances, planetary nebulae.
\end{keywords}

\section{Introduction}
Our Universe has been evolving for 13.8\,Gyr. Over these years many
stars formed and ended their lives enriching the
interstellar medium (ISM), and in consequence enriching the Universe
\citep{planck14}. Very few elements have been around since the
beginning, formed by the Big Bang nucleosynthesis (BBN).  BBN is
responsible for the formation of most of the helium isotope ($^4$He)
in the Universe, along with small amounts of deuterium (D), the helium
isotope ($^3$He), and a very small amount of the lithium isotope
($^7$Li) \citep{BBN48}.  

The predicted abundance (by number) of $^3$He (relative to H) formed
by the BBN (just after a few minutes after the Big Bang) is
1.0$\times$10$^{-5}$ \citep{karakas2014}. This abundance depends only
on the parameter of the current density of baryonic matter.  The
present interstellar $^3$He abundance, as per all the light elements,
comes from a combination of BBN and stellar nucleosynthesis
\citep{wilson94}. H\,{\sc ii} regions are young objects compared with
the age of the Universe, and represent zero-age objects. Their $^3$He
abundance is the result of 13.8\,Gyr of Galactic chemical
evolution. In between is the Solar System, which traces abundances at
the time of its formation, 4.6\,Gyr ago.

Observed values in pre-solar material \citep{geiss93} and the ISM
\citep{gloecker96} imply that
$^3$He/H$=$(2.4$\pm$0.7)$\times$10$^{-5}$.  These values from the
ISM and pre-solar material are approximately twice of the BBN,
implying that the $^3$He abundance has increased a little in the last
13.8\,Gyr.  On the other hand any hydrogen-burning zone of a star
which is not too hot ($>$7$\times$10$^6$\,K ) will produce $^3$He via
the {\it p-p} chain, implying that stars with masses
$<$2.5\,M$_{\odot}$ are net producers of $^3$He. For these stars, {\it
  p-p} burning is rapid enough to produce D {\it in situ}, and enable
the production of $^3$He (D+ p $\rightarrow$ $^3$He + $\gamma$).
Stellar evolution models indeed predict the formation of $^3$He in
significant amounts by stars of 1--2.5\,M$_{\odot}$, with an abundance
of $^3$He/H $> 10^{-4}$ \citep{bania10}, which would have raised the
current $^3$He abundance to $^3$He/H$\sim5\times10^{-5}$,
substantially higher than observed \citep{karakas2014}.

\indent \citet{galli95} presented ``The $^3$He Problem". According to
standard models of stellar nucleosynthesis there should be a $^3$He/H
abundance gradient in the Galactic Disk and the proto-solar $^3$He/H value should
be less than what is found in the present ISM.  Observations of the  $^3$He
abundance in H\,{\sc ii} regions show almost no enrichment above the
BBN value \citep{rood79,bania07}. For the $^3$He problem to be solved, the vast majority of
low-mass stars should fail to enrich the ISM.  One suggestion to solve
this problem is by adding extra mixing in the red giant branch (RGB)
stage. This extra mixing adds to the standard first dredge-up to
modify the surface abundances. \citet{eggleton06} estimate that
while 90\% of the $^3$He is destroyed in 1\,M$_{\odot}$ stars, only
40-60\% is destroyed in a 2\,M$_{\odot}$ star model, depending on the
speed of mixing.

The abundance of $^3$He can only be derived from the
hyperfine transition at the rest frequency of 8.665\,GHz. Detecting
$^3$He$^+$ in PNe challenges the sensitivity limits of all existing
radio telescopes.  \citet{bania10} observed a sample of 12
PNe, $^3$He was detected in only 2 of them. NGC 3242 was observed
with Effelsberg a 100m dish from the Max Planck Institute for Radio
Astronomy (MPIfR) and the National Radio Astronomy Observatory (NRAO)
140-foot telescope, however the observations are inconsistent with
each other. For the case of J320, $^3$He was detected at a 4$\sigma$
level with the NRAO Very Large Array ({\it VLA}). Composite $^3$He$^+$
average spectrum for 6 PNe (NGC 3242, NGC 6543, NGC 6720, NGC 7009,
NGC 7662, and IC 289), using Effelsberg, Arecibo and {\it GBT}
observations, consistently show $^3$He$^+$ emission at the $\sim$1\,mK
level. \citet{me13} observed 3 PNe (IC\,418, NGC 6572,
and NGC 7009) using the {\it VLA}, but no detections were made, and
only upper limits were estimated. 

In this letter we report a 5.7$\sigma$ detection of the $^3$He$^+$ line in the PN IC\,418, using the
National Aeronautics and Space Administration (NASA) DSS-63 antenna of
Robledo de Chavela, Spain. In Sect.~2 the observations are described,
while the results are presented in Sect.~3. The letter ends with a
discussion about the derived abundance of $^3$He$^+$, as well as the
implications to the $^3$He problem.

\section{Observations}
We used the DSS-63 antenna at the Madrid Deep Space Communications
Complex (MDSCC), in Robledo de Chavela, Spain. MDSCC is part of the
NASA's Deep Space Network; the observations were performed under the
``Host Country Radio Astronomy'' program.  The antenna has a diameter
of 70\,m, which results in an angular resolution (half power beam
width) of 115\arcsec, and a sensitivity of 1.25\,Jy/K at 8.6\,GHz. 

We observed the position (RA, Dec)$_{J2000}$ = (05:27:28.2,
-12:41:50), corresponding to the center of IC\,418. The observations
were carried out in different sessions between December 11, 2014 and
March 24, 2015. The observing mode was position switching, which was
done every 2 minutes, with the reference position at 12\arcmin\ from
the source in azimuth. Total on-source integration time was 1090
minutes (i.e., more than 18 hours). Data were
corrected for atmospheric attenuation (all opacities measured were
between 0.02 and 0.04) and elevation-dependent gain, based on a
previously measured gain curve \citep{garcia09}.

We suffered a number of radio frequency interferences (RFIs) at fixed
frequencies (8.330, 8.560, and 8.600\,GHz). Fortunately we did not suffered from RFIs close to the $^3$He$^+$ line.
Furthermore, the bandpass is not totally uniform, which
results in significantly higher system temperatures ($T_{\rm sys}$) at
higher frequencies.  While at the frequency of the H92$\alpha$ line
(8.309\,GHz) $T_{\rm sys}$ is only 22\,K, at the frequency of the
$^3$He$^+$ line it is 210\,K, which results in $rms$ noise
(1-sigma) increasing from 0.3 to 2.3\,mK.

We used the new wideband backend \citep{riz12}, which provides a
frequency resolution of 183\,kHz ($\approx 6$ km\,s$^{-1}$ at
8.5\,GHz) and an instantaneous bandwidth of 1.5\,GHz; we therefore
benefit from the totally usable bandwidth provided by the receiver,
which goes approximately from 8.2 to 9.0\,GHz.

Data was processed using CLASS, a part of the GILDAS
software\footnote{GILDAS is a radio astronomy software developed by
  IRAM. See {\tt http://www.iram.fr/IRAMFR/GILDAS}/.}. We subtracted
the baselines around the detected radio recombination lines (RRLs) and
around the $^3$He$^+$ line to obtain an averaged spectra of every
observing day. After that, spectra from all days have been combined in
a single, final spectrum.

\section{Results}
The Spirograph Nebula (IC\,418, G215.2-24.2) has an
elliptical ring shape, with a major axis of 14$\arcsec$ and a minor
axis of 11$\arcsec$ \citep{ramos12}. It is surrounded by a low-level
ionized halo, which is enshrouded in a neutral envelope with an
angular size of about 2$\arcmin$ \citep{taylor87,taylor89}. The
ionized mass of the nebula is estimated at 0.06\,M$_\odot$, and the
mass of the progenitor at 1.7$\pm$0.3\,M$_\odot$
\citep{morisset09}. 

\begin{figure}
\centering
\includegraphics[width=\columnwidth]{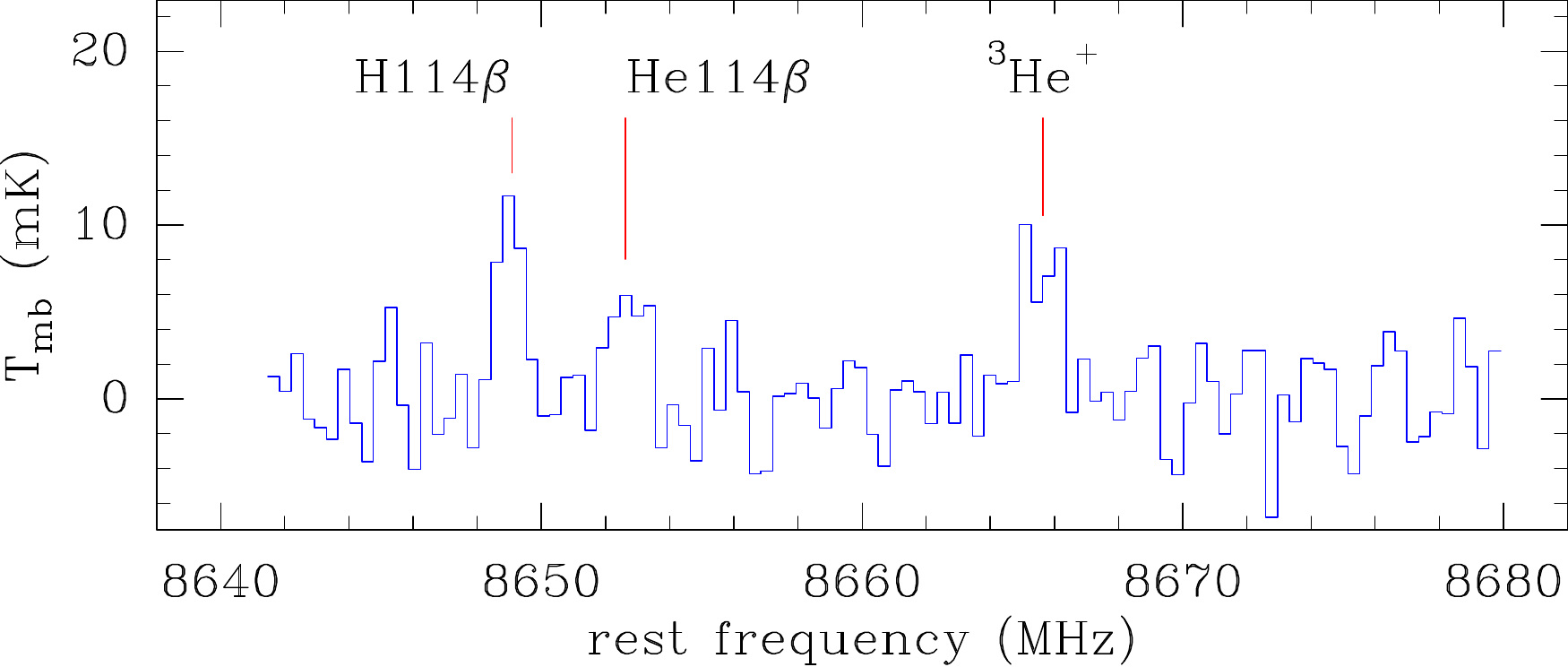}
\caption[Line images1]{DSS-63 observations of the PN IC\,418, showing part of the observed spectrum in a frequency range close to the $^3$He$^+$ line. This line and two close RRLs (H114$\beta$ and He114$\beta$) are indicated. 
Abscissa is the rest frequency for a LSR velocity of 29~km~s$^{-1}$. Data have been smoothed to a a resolution of 12.7~km~s$^{-1}$ (2 channels).}
\label{casa}
\end{figure}

The result of the observations are presented in two figures. 
In Fig. \ref{casa}, the first detection of the $^3$He$^+$ line in IC418 is presented. In 
this figure, a fraction of the spectra (approximately 40 MHz width) is depicted. 
The $^3$He$^+$ line is clearly detected, together with the close RRLs, H114$\beta$ and He114$\beta$. Although the intensity 
of the He114$\beta$ line is quite large compared to the corresponding H line, 
the three lines are centred at the same radial velocity, which reinforces the 
reliability of the detections. 

\begin{figure}
\centering
\includegraphics[width=\columnwidth, height=14.6cm]{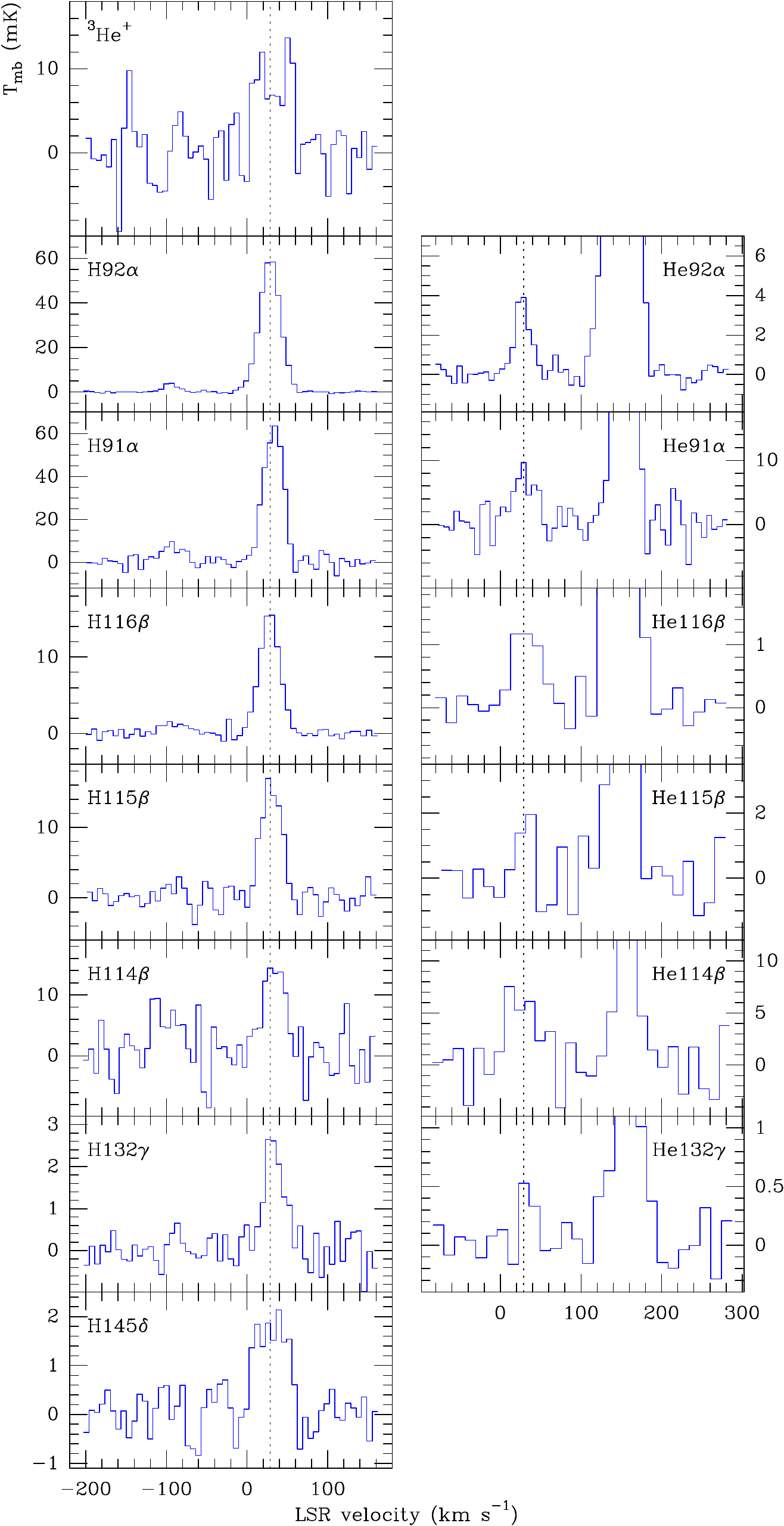}
\caption[Line images1]{DSS-63 observations of the $^3$He$^+$ line and
  the RRLs detected in IC\,418. Lines are indicated at the top left 
corner of each panel, $^3$He$^+$ and the H RRLs on the 
left, and the He RRLs on the right. The full resolution is used, with a channel width of
  6.3\,km\,s$^{-1}$ at 8.66\,GHz except for H132$\gamma$ and the He RRLs that have been 
smoothed to a resolution of 12.7~km~s$^{-1}$ (2 channels) to improve the 
visibility. To ease the comparison, a dashed line at 29~km~s$^{-1}$
  is over-plotted.}
\label{casa1}
\end{figure}

A number of other RRLs have also been observed, as indicated in the Fig. \ref{casa1}. All the lines: H92$\alpha$, H91$\alpha$,
H116$\beta$, H115$\beta$, H114$\beta$, H132$\gamma$, H145$\delta$, and
their He counterparts are centred at 29~km~s$^{-1}$ and display similar line-widths. A Gaussian fitting to the 
detected RRLs are presented in Table 1.

Using the RRLs we analysed the local thermal equilibrium (LTE) and non-local thermal equilibrium (NLTE) conditions for IC\,418 following \cite{brock77} and \cite{gordon09}. We found that LTE and NLTE conditions behave very similarly, NLTE conditions deviate less than 2.5\% from LTE for the H132$\gamma$ and H145$\delta$ lines. When comparing these LTE and NLTE values with the IC\,418 observed ones, we found a small deviation of around 3mK, also seen in the H132$\gamma$ and H145$\delta$ lines. We interpreted this deviation as an instrumental error. For a conservative approach we adopted this error in the measurement of the $^3$He$^+$ line. For a safe LTE analysis, it is optimal to measure the continuum emission at the observed frequency with the same instrument used for the RRLs. Unfortunately, we were not able to do so, and have to infer the continuum. From \citet{me09}, the radio continuum flux density at 8.66\,GHz is 400\,mJy, scaling as $\nu^{-0.1}$ from $S_{\rm 5GHz}=425\pm 5$\,mJy. 

The $^3$He$^+$ column density can be obtained using:
\begin{equation} 
N(^3\rm{He}^+) =\frac{\it{g_l} +\it{g_u}}{\it{g_u}}\frac{8\pi \it{k}\nu^2}
{\it{hc^3\rm{A}_{\it{ul}}}}\int{{\it{T_B}}(\it{v})d\it{v}},
\end{equation}
\noindent where $g_u=$1 is the {\it g} value or magnetic moment of the
upper state, $g_l=$3 is the {\it g} value or magnetic moment of the
lower state, $A_{ul}=$1.95436$\times$10$^{-12}$s$^{-1}$
\citep{gould94}, and $T_B(v)$ is the brightness temperature profile of
the line.

In order to measure the intensity of the $^3$He$^+$ line, we used the task called tdv (from the CLASS
software) which integrates the intensity of the
line over the velocity range.

The flat or double-peaked line profile of the $^3$He$^+$ line differs
from the gaussian profiles of the H and He recombination lines. The
full width at zero maximum of the lines agree, but the outer peaks of
the $^3$He$^+$ line are lacking from the recombination lines. We
therefore separated the inner and outer parts of the line profiles
through a triple gaussian fit. The inner component, with a FWHM of
18\,km\,s$^{-1}$, consistent with the recombination lines, is used
as a lower value to the line intensity. The full integrated intensity
is used as the upper value. We then used the beam filling factor (bff)
to derive the observed beam-averaged brightness temperature $T_{mb}$ \citep{garcia09}.

The number density of $^3$He$^+$ atoms, $n$($^3$He$^+$), can
be obtained by dividing the column density $N$($^3$He$^+$) by the
averaged optical path, $<\Delta{s}>$, through the source. Representing
a PN as a homogeneous sphere of radius $R=\theta_sD$, where $D$ is the
distance, the optical path at angle from the centre $\theta$ is
$\Delta{s}(\theta)=2\sqrt{R^2-(\theta D)^2}$, and the optical path
averaged over the source becomes $<\Delta{s}>=\pi R/2 = \pi
\theta_sD/2$.

The expression for  $n$($^3$He$^+$) is then
\begin{equation} 
\label{density}
\frac{n(^3\rm{He}^+)}{\rm{cm^{-3}}}= 44.6\left(\frac{T_{mb}\,\Delta{\it{v}}}{\rm{mK\,km\,s^{-1}}}\right)\left(\frac{D}{\rm{kpc}}\right)^{-1}\left(\frac{\theta_{s}}{{\arcsec}}\right)^{-3}\left(\frac{\theta_{b}}{57.5\arcsec}\right)^2\left(1+\frac{\theta_{s}^{2}}{\theta_{b}^{2}}\right)
\end{equation}

Table \ref{abun} presents the line peak ($T_{mb}$), the
integrated intensity, and the width ($\Delta v$) of all the lines
observed. For the $^3$He$^+$ we integrated the intensity of the
line; for the RRLs we fitted a gaussian to estimate their intensities.
The intensity ratios of the Hn$\alpha$ and Hn$\beta$ are consistent
with model predictions to within 10\%, for an emission measure of
EM$\sim 7 \times 10^6\,\rm pc\,cm^{-6}$ \citep{hjellming1969}. The
line-to-continuum ratio $\sim$4.5\%, is also consistent with this EM.
This gives confidence in the bandpass and flux calibration.

To calculate the fractional $^3$He abundance we 
divide by the H$^+$ density. The  H$^+$ density was
modelled using Cloudy\_3D \citep{morisset06}. 
In this Cloudy\_3D model for IC418 the nebula is an ellipsoid with a small eccentricity. Because of the small eccentricity, a spherical version of the nebula was also tested, giving similar results, with the caveat that the radial density profile is not uniform, two shells were needed. From that model, a mean H+ density can be derived. The value presented in
Table \ref{parameters} was taken from \citet{morisset09}.  The
distance  used was 1.3kpc \citep{me09}.

\begin{table}
\caption[IC\,418 lines]{$^3$He$^+$ line and radio recombination lines 
detected in the PN IC\,418.}
\label{abun}
\centering
\begin{tabular}{llll} 
\hline\hline 
Line & $T_{mb}$  & Area    & $\Delta v$   \\
        & (mK)    & (mK\,km\,s$^{-1}$) & (km\,s$^{-1}$)\\
\hline\hline
$^3$He$^+$ & 10.6$\pm$3.0 & 369.8$\pm$64.7 & -- \\
H92$\alpha$ & 59.8$\pm$0.6& 1925.8$\pm$15.0 & 30.2$\pm$0.2\\
He92$\alpha$ & 3.6$\pm$0.5 & 72.0$\pm$8.4 & 21.9$\pm$1.9\\
H91$\alpha$ & 64.3$\pm$3.0 & 1884.1$\pm$59.4 & 27.5$\pm$0.9\\
He91$\alpha$ & 7.1$\pm$3.3 & 210.7$\pm$53.5 & 28.3$\pm$6.6\\
H116$\beta$ & 16.0$\pm$1.3 & 507.3$\pm$26.5 & 29.6$\pm$1.9\\
He116$\beta$ & 1.3$\pm$0.5 & 59.2$\pm$29.4& 40.7$\pm$23.4\\   
H115$\beta$ & 16.7$\pm$1.6 & 507.1$\pm$32.3 &  28.4$\pm$1.9\\
He115$\beta$ & 2.1$\pm$1.4 & 48$\pm$23 &  32.3$\pm$7.1\\
H114$\beta$ & 14.5$\pm$3.7 & 470.5$\pm$77.3 & 30.4$\pm$5.5 \\
He114$\beta$ & 5.5$\pm$4.0 & 121.5$\pm$85.2 & 36.0$\pm$12.8\\
H132$\gamma$ & 2.6$\pm$0.5& 78.9$\pm$10.8 & 28.3$\pm$4.8 \\
He132$\gamma$  & 0.6$\pm$0.3 & 9.1$\pm$4.8 &12.5$\pm$7.7\\  
H145$\delta$ & 2.0$\pm$0.5 & 87.9$\pm$11.8 & 41.3$\pm$5.3\\
 \hline
   \end{tabular} 
\end{table}

\begin{table}
\caption[IC\,418 abundance]{The $^3$He
  abundance in the PN IC\,418. The upper row uses the central
  component of the $^3$He line only; the bottom row uses the full
  integrated line profile. These are used as lower and upper values
  respectively. An ICF (He$^+$) of 1.45 was used.}
\label{parameters}
\centering
\begin{tabular}{cccccc} 
\hline
\hline 
D & $\theta_{s}$ &  T$_{mb}\Delta v$ &  $n$(H$^+$) &  $n$($^3$He$^+$) & $^3$He/H \\
(kpc) & ($\prime\prime$) & (mK\,km\,s$^{-1}$) & (cm$^{-3}$) & (cm$^{-3}$) & (10$^{-3}$)\\
\hline\hline
1.3$\pm$0.4 & 7 & 109.1$\pm$42.3& 9.3$\times$10$^3$ & 11.1$\pm$4.9 & 1.74$\pm$0.8 \\	
1.3$\pm$0.4 & 7 & 369.8$\pm$64.7 & 9.3$\times$10$^3$ & 37.5$\pm$10.5 & 5.8$\pm$1.7 \\	
 \hline
   \end{tabular} 
\end{table}

From the detailed model of IC\,418 described by
\citet{morisset09}, the Ionisation Correction Factor (ICF) of He$^+$
is 1.45 (taking into account the presence of an He$^0$ region). This
yields an abundance of $^3$He/H=1.74$\pm$0.8$\times$10$^{-3}$ for the
lower limit, and 5.8$\pm$1.7$\times$10$^{-3}$ for the upper limit.

The recombination lines allow us to estimate the total He/H ratio
\citep{roelfsema1987}. We calculated the ratio of the line strength
integrated over the profile, for the five available line couples
(He91$\alpha$/H91$\alpha$, etc). The ratios are multiplied by the ICF
for helium of 1.45, and divided by 1.07 to account for the slightly
different radiative recombination rates of hydrogen and
helium. He92$\alpha$ is a factor of three fainter than expected and we
suspect a fitting error, while He114$\beta$ is too faint for a
reliable value. The remainder give He/H$=0.15 \pm 0.03$.  $^3$He
accounts for 4\%\ of the helium in IC\,418.

\section{Discussion}
\subsection{Origin of the $^3$He$^+$ emission}
Three aspects of the current data needs to be considered. First, the
derived $^3$He$^+$ abundance is well above model expectations. Second,
the upper value for the $^3$He$^+$ abundance is only a little lower
than the earlier 3-$\sigma$ VLA upper limit. Third, although the full
width of the $^3$He$^+$ line is consistent with the optical expansion
velocity of IC\,418, the profile differs from that of the
recombination lines, peaking at the outermost velocities.

A double-peaked profile could arise from an expanding detached shell,
which is larger than the beam. In this case, the outer components of
the profile, which differ from the recombination lines, come from this
large region, whilst the central part of the profile arises from the
inner, ionized nebula. \citet{Balser1999} also proposed a contribution
from a large, low density halo. The emission in the $^3$He$^+$
hyperfine line scales with $\int n\, dr$ (column density), and that of
recombination line with $\int n^2\,dr$. Therefore, a low density but
high mass halo could explain the difference in profiles.

\begin{figure}
\centering
\includegraphics[width=\columnwidth, height=6.5cm]{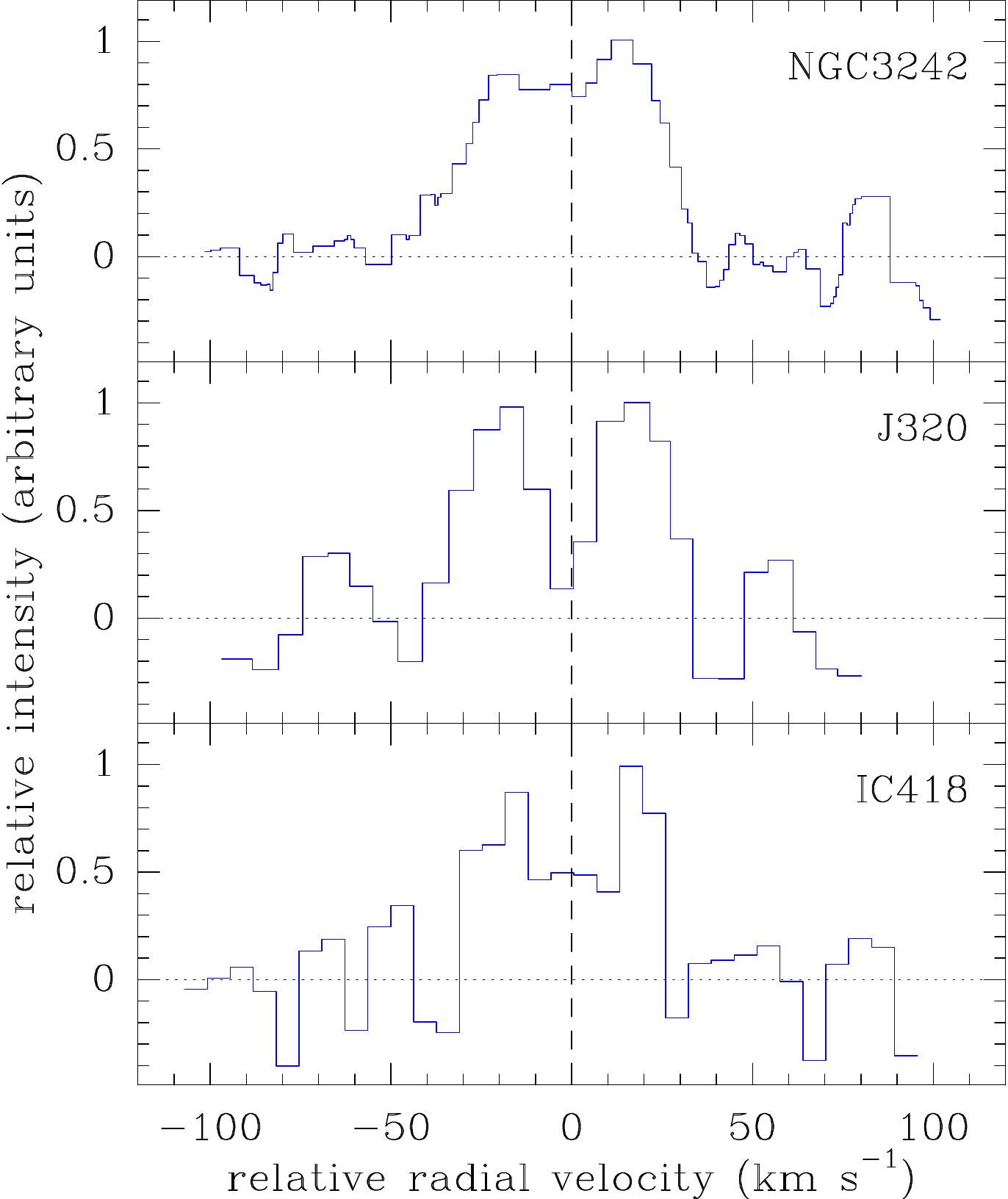}
\caption{All the $^3$He$^+$ line detected in planetary nebulae. The sources are NGC\,3242 
(Balser et al. 1999), J320 (Balser et al. 2006), and IC\,418 (this work). In order 
to facilitate the comparison, the abscissa is presented as radial velocities 
relative the LSR systemic velocity of each source (dashed-line), and the intensity scale as 
fractions of the peak intensity of each source.}
\label{prevpn}
\end{figure}

Two other PNe have reported $^3$He$^+$ detections: J320
\citet{Balser2006} and NGC 3242
\citet{balser97,Balser1999,Balser2006}. Both have double peaked
profiles, similar to IC\,418 (Fig. \ref{prevpn}). Both objects have
haloes, in the case of NGC\,3242 possibly as large as 18 by 24\,arcmin
diameter \citet{Bond1981}. Whether helium in such a halo could be
photo-ionized by the star is not clear.  NGC\,3242 has an inner shell
and an outer elliptical envelope, but the inner shell is optically
thick to He$^+$ ionizing photons \citep{Ruiz2011}. 
%IC\,418 is also ionization-bounded.

If part of the line arises from a halo, then the fractional abundance
should be calculated excluding this part. The reason for this is that
the ionized mass derived from recombination or free-free emission,
as used in the calculations, is not sensitive to such haloes.  The total
mass of ionized haloes is very poorly constrained. We therefore
consider the low value for the $^3$He/H ratio for IC\,418, which
was derived for the line component coinciding with the recombination
lines, as more likely.

\subsection{Zeeman splitting}
Zeeman splitting in the $^3$He$^+$ line is similar to that of the HI
line. The line frequency moves by $1.399\,\rm kHz mG^{-1}$, or
$0.48\,\rm km\,s^{-1}\,mG^{-1}$. The two $\sigma$ components are
separated by twice this. For a typical field strength within PNe of a
few mG \citet{Sabin2007}, the line shifts in frequency by one to a
few km\,s$^{-1}$. This is insufficient to explain the difference in
line profile. 

\subsection{Hyperfine maser}
\cite{deguchi85} show that the spin temperature of $^3$He$^+$ can go
negative only if [He$^{++}$/He$^+$]$>>0$. The inversion in this case
is driven by $^4$He Ly$\alpha$ absorption. This effect requires very
high excitation. Some PNe  do contain regions of such high ionization.

The outer peaks of the $^3$He$^+$ profiles could show the effect of
amplification. The inner peak would give the actual abundance,
unaffected by non-thermal amplification.  The preferred value for the
$^3$He$^+$ fractional abundance is again the lower value derived
above.

\subsection{Stellar evolution}
\begin{figure}
\includegraphics[width=\columnwidth, height=7.3cm]{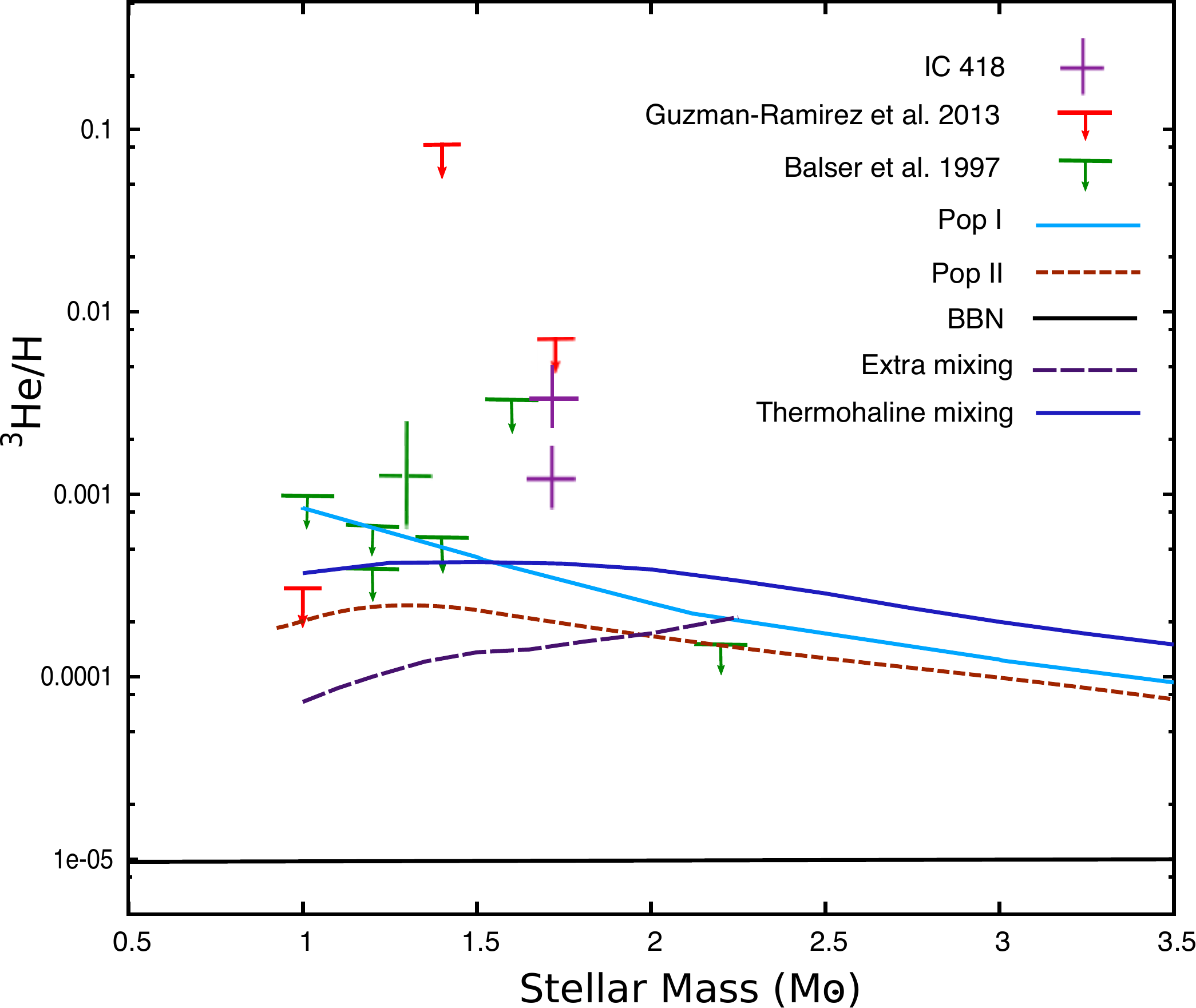}
\caption[$^3$He/H abundance]{Abundances of $^3$He (relative to H) for
  the PN IC\,418 (both purple crosses, for upper and lower estimate),
  the size of the bars correspond to the uncertainties. Red arrows are
  the upper limits from \citet{me13}. The green cross
  represents J320 and the green arrows represent the sample of 6 PNe from
  \citet{balser97}. The curves for Pop I (light-blue line) and II (red
  dashed-line) show the standard abundance of $^3$He taken from
  \citet{weiss96}. The purple dashed-line labelled Extra mixing
  represents the results of models including deep mixing by
  \citet{boothroyd99}, and the blue line shows the stellar models using
  thermohaline mixing \citep{charbonnel07}. The black line is the primordial value of
  $^3$He/H from the BBN \citep{karakas2014}.}
\label{abundances}
\end{figure}

The contribution of PNe to the $^3$He
abundance is crucial for understanding the Galactic chemical evolution. 
Figure \ref{abundances} shows the $^3$He abundances of the PN IC\,418
(both purple crosses).  For comparison the upper limits calculated from
\citet{me13} observations are presented using the red
arrows. \citet{balser97} observations are shown in green: the 
cross represents J320 and the arrows are the upper limits, where
the mass estimates are from \citet{galli97}.  The stellar evolution
models are also presented in Figure \ref{abundances}. 

Even for the lower value for IC\,418, which we consider as more
likely, the abundance is above the model calculations for any
mass. This present problems for models which invoke deep mixing (purple dashed-line)
\citet{boothroyd99}. The STARS models (blue line) with extra mixing via thermohaline convection by \citet{charbonnel07} come closest. 

\section{Conclusions}

We have detected the $^3$He$^+$ line towards the PN IC\,418 using the
DSS-63 antenna. We derive $n(^3{\rm He}^+)$=11$\pm$5--38$\pm$11cm$^{-3}$, corresponding to $^3$He/H=1.7$\pm$0.8--5.8$\pm$1.7$\times$10$^{-3}$. We have some
preference for the lower range, which is derived using the central
component of the line only. The outer component may arise in a
extended low density halo.

The values exceed prediction from stellar evolution models, especially
those invoking deep mixing. The lower limit is a factor of 2 above
the thermohaline model and we may be approaching consistency with it. However, the large amounts of $^3$He produced in these models
is at odds with the abundance of $^3$He observed in the ISM and the
Solar System ($\sim$10$^{-5}$). The $^3$He problem lingers.

\section*{Acknowledgments}
This work is based on observations made with the DSS-63 antenna at the
MDSCC, under the ``Host Country Radio Astronomy'' program. LGR thanks Arturo Manchado for bringing the DSS-63 antenna to her
attention. LGR is co-funded under the Marie Curie Actions of the
European Commission (FP7-COFUND). CM acknowledges the UNAM project
PAPIIT-107215

\newcommand{\mockalph}[1]{}

\bibliographystyle{mn2e}
\addcontentsline{toc}{chapter}{\bibname}
\bibliography{biblioThesis}
\bsp

\label{lastpage}

\end{document}